\documentclass[11pt,twoside]{article}
\usepackage{asp2004}
\usepackage{graphics, epsfig}
\begin{document}
\title{Mass and angular momentum loss during RLOF in Algols}
\author{W. Van Rensbergen, C. De Loore $\&$ D. Vanbeveren}
\affil{Astrophysical Institute, Vrije Universiteit Brussel, Pleinlaan 2, B-1050 Brussels, Belgium}

\begin{abstract}
We present a set of evolutionary computations for binaries with a B-type primary at birth. Some liberal 
computations including loss of mass and angular momentum during binary evolution are added to an extensive
grid of conservative calculations. Our computations are compared statistically to the observed distributions
of orbital periods and mass ratios of Algols. Conservative Roche Lobe Over Flow (RLOF) reproduces the
observed distribution of orbital periods decently but fails to explain the observed mass ratios in the range
$\in$ $\lbrack$0.4-1$\rbrack$. In order to obtain a better fit the binaries have to lose a significant amount
of matter, without transferring too much angular momentum.
\end{abstract}
\thispagestyle{plain}

\section{Introduction}

\citet{Eggleton} introduced the denomination "liberal" to make a distinction between binary evolution with 
mass and angular momentum loss and the conservative case where no mass and consequently no angular momentum
leave the system. \citet{Refsdal et al.} showed that the binary AS Eri is the result of liberal binary
evolution; the amounts of mass and angular momentum lost by the system are however uncertain. \citet{Sarna}
showed that only 60 $ \%$ of the mass lost by the loser of $ \beta$ Per was captured by the gainer and that
30 $ \%$ of the initial angular momentum was lost during RLOF. Hence it is clear that the liberal
evolutionary scenario is  important for binary evolution calculations. However the amount of mass and angular
momentum that has to be removed from interacting systems is far from obvious.

With the Brussels simultaneous evolution code (a description is given in "The Brightest Binaries", 
\citet{Vanbeveren et al.}) we calculated a representative grid of conservative evolution of binaries with a B
type primary at birth. Application of the criterion of  \citet{Peters} allows then to determine for each of
the evolutionary sequences the beginning and ending of various Algol-stages. This criterion states that in
the semi-detached system:
\begin{flushleft}
$\bullet$~~~The less massive star fills its Roche lobe (RL)\\ 
$\bullet$~~~The most massive star does not fill its RL and is still on the main sequence\\ 
$\bullet$~~~The less massive star is the coolest, the faintest and the largest\\  
\end{flushleft}

The grid allows to determine an expected distribution of orbital periods and mass ratios of Algols. These 
can be compared to the well established observed distributions (\citet{Budding et al.}).

Considering the examples mentioned above (\citet{Refsdal et al.}, \citet{Sarna}) it may be expected that 
the match between the observations and the conservative results are far from satisfactory. Therefore we add
liberal calculations to our conservative library. The detailed evolutionary tracks can be found in
\verb"http://www.vub.ac.be/astrofys/"

\section{Details of the conservative calculation}

\begin{figure}[h]
\begin{center}
\epsfig{file=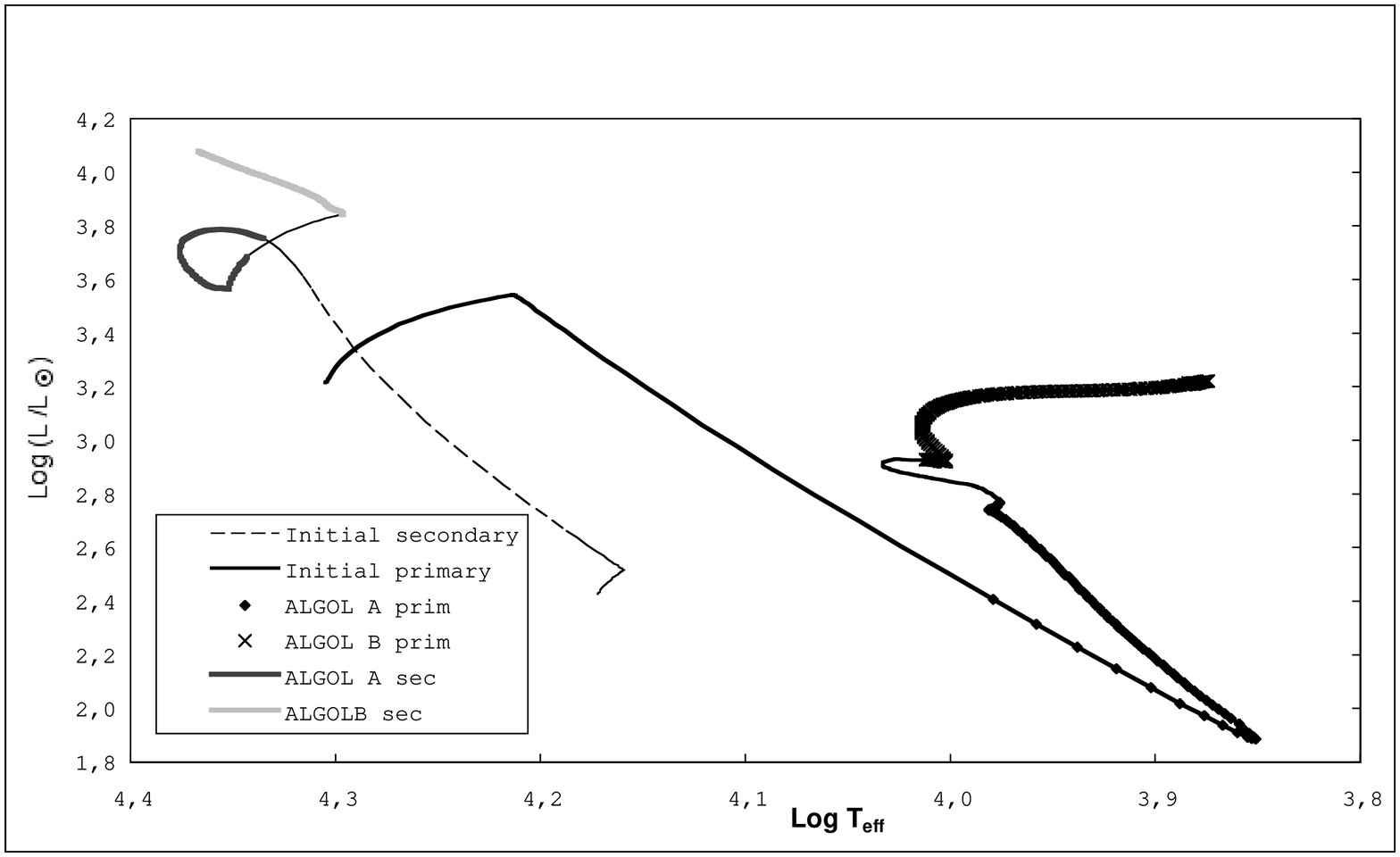, height=6cm,width=12cm}
\caption{\it Conservative evolution of a (7+4.2)$M_{\odot}$ binary with an intial period of 2.5 d. The 
system remains Algol A for $\approx$ 20 million years before it is Algol B for $\approx$ 1.5 million years.}
\label{fig_fig1}
\end{center}
\end{figure}

Our grid contains binaries with sufficiently small initial periods to lead to Case A RLOF (RLOF A): i.e. 
during H core burning of the donor (initially primary star that becomes the less massive after
Algol-ignition). The track across the HRD is traced for every system in our grid. A binary lives its era of
"Algolism" (\citet{De Loore et al.}) when it obeys the criterion of  \citet{Peters}.  Every binary shows its
Algol A (Algol during H core burning) aspect for some time during RLOF A. The drastical change of the mass
ratio and the orbital period during this process can be followed in detail for every system in our grid. It
happens frequently that RLOF A is succeeded by Case B RLOF (RLOF B): i.e. during H shell burning of the
donor. Systems that have sufficiently large initial orbital periods so that RLOF A does not occur will also
show a short living Algol B (Algol during H shell burning) appearance. These systems have been considered in
a previous paper (\citet{Van Rensbergen02}). Figure \ref{fig_fig1} shows a typical example of a case AB. The
(7+4.2)$M_{\odot}$ binary with an intial period of 2.5 d starts as Algol A with q=1 and P=2.06 d. It ends
this Algol A stage after some 20 million years with q=0.27 and P=6.87 d. From these inital values the system
remains an Algol B for some 1.5 million years during H shell burning of the donor. The system eventually
evolves into a long periodic (P${>}$50 d) Algol B with q $\approx$ 0.1. The grid contains some 240 more
evolutionary tracks.

\section{Conservative simululation}

From our grid of conservative calculations we made a simulation of the distribution of orbital periods and 
mass ratios of Algols using a Monte Carlo algorithm. As starting conditions we selected:  

\begin{flushleft}
$\bullet$~~~The primaries IMF of \citet{Salpeter}: $\zeta$(M)$\div$~$M^{-2.35}$\\ 
$\bullet$~~~an initial distrubution of orbital periods from  \citet{Popova et al.}: $\Pi$(P)$\div$$\frac{1}{P}$\\ 
$\bullet$~~~an initial mass ratio distribution as derived by \citet{Van Rensbergen01} from the non-evolved 
systems of the catalogue of Spectroscopic Binaries shown on \verb"http://sb9.astro.ulb.ac.be/" \\  
\end{flushleft}

The distribution of the initial mass ratio q = $\frac{M_{2}^{0}}{M_{1}^{0}}$ obeys relation (1):

\begin{equation}
\Psi(q)\div~(1+q)^{-\alpha}~;~\alpha= 3.37~for~early~B~\&~1.47~for~late~B~primaries
\end{equation} 

\begin{figure}[h]
\begin{center}
\epsfig{file=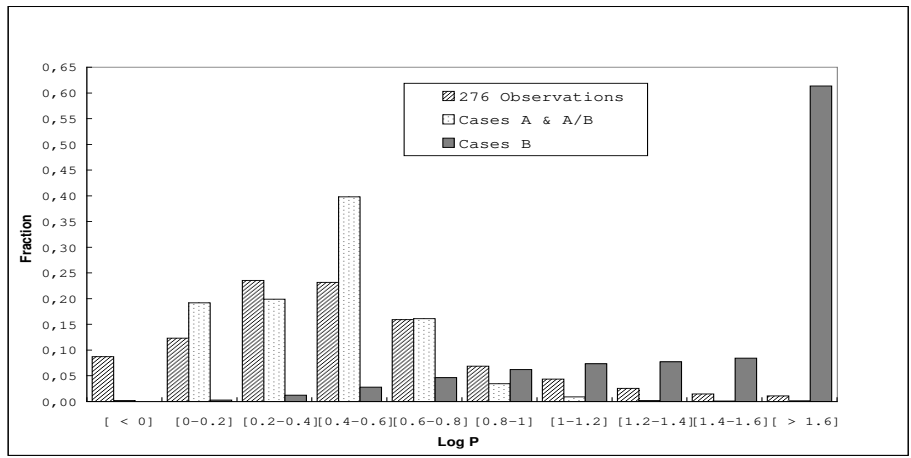, height=6cm,width=12cm}
\caption{\it Observed distribution of the orbital periods of Algols compared to conservative evolution. 
Cases B produce an excessive amount of long periods. Cases A follow the observed distribution better. There
is no doubt that there are far more Algol A than Algol B cases.}
\label{fig_fig2}
\end{center}
\end{figure}

Figure  \ref{fig_fig2} shows the observed distribution of orbital periods of Algols. Here we find more A 
than B-cases, because the Algol A-phase goes on for a fraction of the nuclear time scale, whereas the Algol
B-phase lasts only for a fraction of the much shorter Kelvin-Helmholtz time scale. RLOF A produces Algols
that follow the observed distribution rather well. Since the periods of the Algol B cases peak towards the
long periods, a contribution of a few $\%$ of B cases to the Algol population will mimic the observed period
distribution well.

\begin{figure}[h]
\begin{center}
\epsfig{file=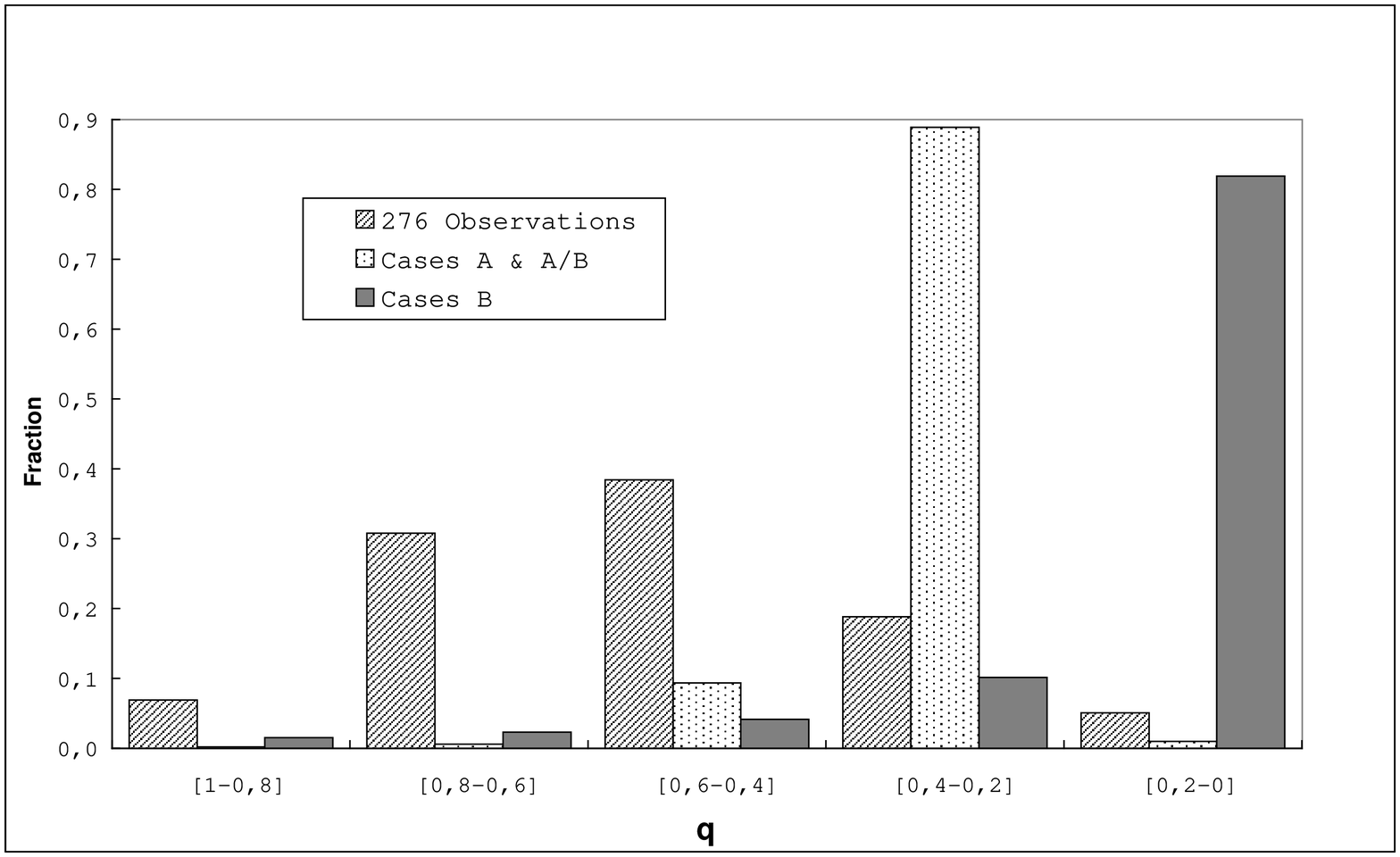, height=6cm,width=12cm}
\caption{\it Observed distribution of mass ratios of Algols compared to conservative evolution. Cases B 
produce more than 80$\%$ Algols with the smallest mass ratio  q $\in$ $\lbrack$0-0.2$\rbrack$. Cases A
produce most of their Algols with q $\in$ $\lbrack$0.2-0.4$\rbrack$. The fact that almost 80$\%$ Algols are
observed with q $\in$ $\lbrack$0.4-1$\rbrack$ excludes conservative evolution as the major channel through
which Algols can be formed.}
\label{fig_fig3}
\end{center}
\end{figure}

Figure  \ref{fig_fig3} shows the observed mass ratio distribution of Algols. The mass ratio q is now defined 
as the observers do: q = $\frac{M_{donor}}{M_{gainer}}$.

The mass ratios of the Algol B cases peak towards the smallest mass ratios, whereas RLOF A produces a 
majority of Algols in the q-bin ${\lbrack}$0.2-0.4${\rbrack}$. As a consequence an admixture of cases A and B
will never reproduce well the observed mass ratio distribution in the q-bins ${\lbrack}$0.4-1${\rbrack}$. We
may conclude that liberal binary evolution is needed to describe the mass ratio distribution of Algols.

\section{Confining the liberal model}

Mass loss is defined by a parameter $\beta$ describing the fraction of the mass lost by the loser that is 
accreted by the gainer: 

\begin{equation}
\dot{M}_{gainer} = - \beta~\dot{M}_{donor}~~~with~~~0~~\leq~~\beta~~\leq~~1
\end{equation} 

Conservative evolution is described with $\beta$ = 1, whereas the liberal case uses values of $\beta$ $<$ 1 
which are not known beforehand. Conservative evolution implies that no angular momentum can be lost by the
system, whereas the amount of loss of angular momentum in the liberal case is also a free parameter if no
physics restricts the assumptions. It is clear that realistic hydrodynamical calculations should learn us the
appropriate choice of the amounts of mass and angular momentum which are lost by a binary at any moment of
the RLOF process. In the mean time we have performed a number of liberal evolutionary calculations and
compared the results with the observations.

Our calculations reveal that the time dependent parameter $\beta$(t) should sufficiently often differ 
drastically from 1. 

For a given value of $\beta$, the amount of angular momentum lost by the system is defined by the position 
of the site where matter is recoiled from the system.  \citet{Soberman et al.} have argued that matter can be
trapped in a Keplerian ring after transit across the second Lagrangian point $L_{2}$. The radius of the ring
is $\eta$ times the semi major axis of the orbit. This Keplerian ring passes outside $L_{2}$ which is located
at $\eta$ $\approx$ 1.25. This yields a minimum value of $\eta$ $\approx$ 1.25. Hydrodynamical calculations
of  \citet{Lubow and Shu} locate the ring at $\eta$ $\approx$ 3. This yields a maximum value of $\eta$
$\approx$ 3. A Keplerian ring located at $\eta$ $\approx$ 2.25 takes away as much angular momentum as  the
co-rotating point $L_{2}$ would do. A value of $\eta$ $\approx$ 2.25 is thus a fair value in the interval
${\lbrack}$1.25-3${\rbrack}$ that can be used to calculate the change of the orbital period of a binary as a
consequence of loss of mass and angular momentum through a ring which rotates with Keplerian velocity around
the center of mass of the system (\citet{Soberman et al.}).

Our calculations reveal that if the system loses angular momentum across points located near 
$\eta$ $\approx$ 2.25 the orbital periods shrink drastically. Most binaries become mergers before they show
Algolism. The obtained distribution of orbital periods of Algols is completely shifted towards the shortest
periods. Since this  conclusion conflicts with observations, the mass that leaves the system carries rather
the angular momentum of points located near $\eta$ $\approx$ 0 than near $\eta$ $\approx$ 2.25.

\begin{figure}[h]
\begin{center}
\epsfig{file=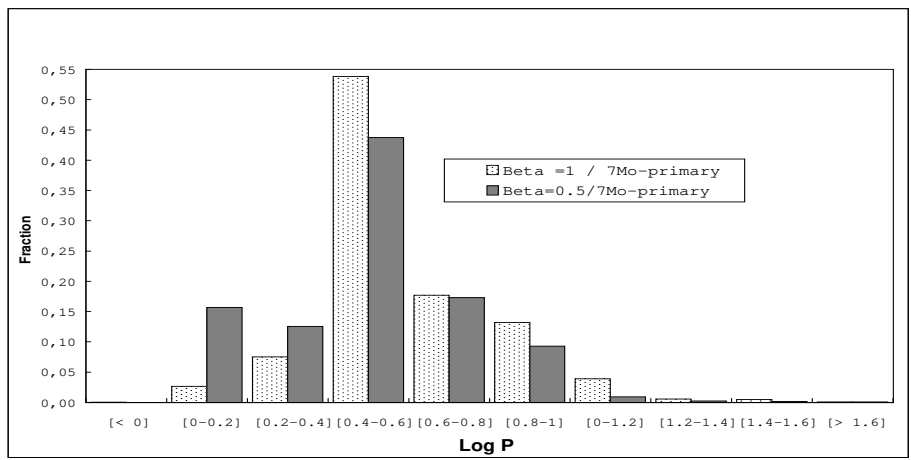, height=6cm,width=12cm}
\caption{\it Liberal evolution with a lot of mass loss ($\beta$=0.5) and a little loss of angular momentum 
leaves the distribution of orbital periods of Algols almost unaltered. This figure shows the result for
Algols issued from RLOF A and a 7 $M_{\odot}$  primary at birth.}
\label{fig_fig4}
\end{center}
\end{figure}

\begin{figure}[h]
\begin{center}
\epsfig{file=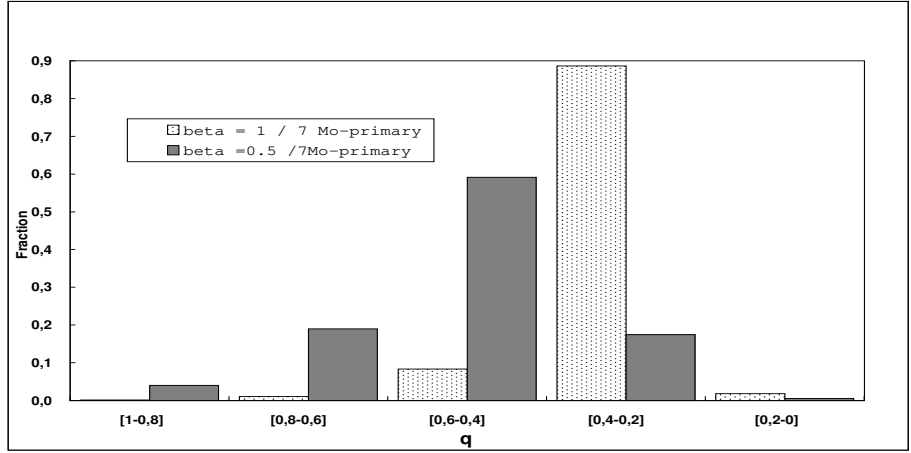, height=6cm,width=12cm}
\caption{\it Liberal evolution with a lot of mass loss ($\beta=0.5$) and a little loss of angular momentum 
changes the mass ratio distribution of Algols drastically. The observed mass ratios in the q-interval
$\lbrack$ 0.4-1 $\rbrack$ which were not produced by conservative evolution are created by this liberal
model. This plot shows the result for Algols issued from RLOF A and a 7$M_{\odot}$ primary at birth.}
\label{fig_fig5}
\end{center}
\end{figure}

For all these (statistically sound but physically not well understood) reasons we performed a number of 
calculations with constant $\beta$ = 0.5 and angular momentum lost at the edge of the gainer. Until now only
the representative cases with a 7$M_{\odot}$ primary at birth and initial periods leading to RLOF A have been
performed. Figure \ref{fig_fig4} shows that the assumption of mass loss through a point near $\eta$ = 0 does
not alter the conservative distribution of the orbital periods of Algols. A conclusion that meets the
observations since the conservative orbital period distribution matched the observations fairly well. Figure
\ref{fig_fig5} shows that our liberal assumption  ($\beta$=0.5, $\eta$ $\approx$ 0) deviates radically from
the conservative case. The q-bins ${\lbrack}$0.4-1${\rbrack}$ which were not populated in the conservative
scenario are now populated properly as required by the observations (see figure  \ref{fig_fig3}).

\begin{quote}

\end{quote}

\end{document}